\documentclass[11pt, aps,superscriptaddress]{revtex4}

\usepackage{latexsym}
\usepackage{amsthm}
\usepackage{amssymb}
\usepackage{amsmath}

\usepackage[all]{xy}
\usepackage{graphicx}
\usepackage{subfigure}

\usepackage{dcolumn} 
\usepackage{bm} 
\usepackage{color}

\SelectTips{eu}{11}

\usepackage[colorlinks=true,citecolor=blue, linkcolor=blue]{hyperref}

\begin{document}

\title{An Efficient Deterministic Quantum Algorithm for the Integer Square-free Decomposition Problem}
\author{Jun Li}

\author{Xinhua Peng}
\email{xhpeng@ustc.edu.cn}

\author{Jiangfeng Du}
\email{djf@ustc.edu.cn}
\affiliation{Hefei National Laboratory for Physical Sciences at Microscale and Department of Modern Physics, University of Science and Technology of China, Hefei, Anhui 230026, People's Republic of China}
\author{Dieter Suter}
\affiliation{Fakult\"{a}t Physik, Technische Universit\"{a}t Dortmund, 44221 Dortmund, Germany}

\begin{abstract}
Quantum computers are known to be qualitatively more powerful than classical computers, but so far only a small number of different algorithms have been discovered that actually use this potential. It would therefore be highly desirable to develop other types of quantum algorithms that widen the range of possible applications. Here we propose an efficient and deterministic quantum algorithm for finding the square-free part of a large integer - a problem for which no
efficient classical algorithm exists.
The algorithm relies on properties of Gauss sums and uses the quantum Fourier transform.
We give an explicit quantum network for the algorithm.
Our algorithm introduces new concepts and methods that have not been used in quantum information processing so far
and may be applicable to a wider class of problems.
\end{abstract}
\maketitle

A fundamental tenet of classical computer science is based on the Church-Turing thesis, which asserts that any practically realizable computational device
can be simulated by a universal computer known as the Turing machine  \cite{T}. However, this hypothesis implicitly relies on the laws of classical physics \cite{L} and was challenged by Feynman  \cite{F} and others who suggested that computational devices behaving according to quantum mechanics could be qualitatively more powerful than classical computers. A first proof of this conjecture was given in 1993 by Bernstein and Vazirani  \cite{BV}. They showed that a quantum mechanical Turing machine is capable of simulating other quantum mechanical systems in polynomial time, an exponential improvement in computational power over the classical Turing machine. Their proof did not give an actual fast quantum algorithm, but in the following year, Peter Shor came up with his famous factoring algorithm \cite{S}, which solves the integer factorization problem in polynomial time, exponentially faster than any known classical algorithms. The essential part of this algorithm is a solution of the order-finding problem, which can be formulated as a hidden subgroup problem (HSP) \cite{NC}. A hidden subgroup problem is like to find out the period of a given periodic function. The structure of the function's periodicity may be so complicated that it can not be easily determined by classical means.
The importance of the HSP is that various instances (eg. Pell's equation, the principal ideal problem, unit group computing) and variants like
the hidden shift problem and hidden nonlinear structures encompass most of the quantum algorithms found so far
that are exponentially faster than their classical counterparts \cite{CD}.
This relatively narrow range of existing fast quantum algorithms shows the urgent need for different types of quantum algorithms
that will make other classes of problems accessible to efficient solutions.

Here we describe such a quantum algorithm that does not fall into the framework of HSP.
It solves two number-theoretical problems in polynomial time, i.e., testing the square-freeness and computing the square-free part of a given integer.
Compared to the known classical algorithms, this provides an exponential increase  in computational efficiency.
While these problems are related to the factorization problem solved by Shor, our algorithm relies on a different approach.
Furthermore, while Shor's algorithm is probabilistic, the algorithm presented here is deterministic and its  computational complexity is lower.

We consider a positive integer $N$ with its unique prime factor decomposition $N = p_1^{{\alpha _1}}p_2^{{\alpha _2}} \cdots p_k^{{\alpha _k}}$ ($p_i$ are primes). $N$ is called square-free if no prime factor occurs more than once, i.e., for all $i$ ($i=1,2,...,k$), ${{\alpha _i}}=0$ or $1$.
An arbitrary positive integer can always be written as
\begin{equation}
N = r \cdot {s^2},
\end{equation}
where $r$ is square-free, and this square-free decomposition is unique.
Thus, usually $r$ and $s^2$ are called the square-free part and the square part of $N$, respectively.
The square-freeness testing problem corresponds to determining whether $s=1$.
An additional problem consists in finding the square-free part $r$ of $N$.
These problems were listed as two unsolved open problems \cite{AM}, since no efficient algorithm is currently known for either of them.
Actually they may be no easier than the general problem of integer factorization \cite{BL}.
It was found  \cite{OU} that the factorization of  $N = p{q^2}$ ($p$, $q$ both prime) is almost as hard as the factorization of $N = pq$. This fact has been used in a proposed digital signature scheme called TSH-ESIGN, which is more efficient than any representative signature scheme such as elliptic curve and RSA based signature \cite{OU}. On the other hand, the square-free part problem appears to be a representative of a larger class of computational problems. As an example, computing the ring of integers of an algebraic number field, one of the main tasks of computational algebraic number theory, reduces to it in deterministic polynomial time \cite{C, L92}.

We now describe an efficient, deterministic quantum algorithm that solves both problems.
It uses the Gauss sum, an important object which has been extensively investigated in mathematics (see supplementary information). Throughout this paper, we will assume that $N$ is an odd integer (the case of even numbers can be trivially reduced to this case).
The Gauss sum is defined as
\begin{equation}
G(a,\chi ) = \sum\limits_{m = 0}^{N - 1} {{\chi _N}(m){e^{2\pi iam/N}}} ,  \nonumber
\end{equation}
where $a$ is an integer and the function ${{\chi _N}(m)}$ represents the Jacobi symbol of $m$ relative to $N$ \cite{IR}.

The evaluation of the Gauss sum is closely related to the square-freeness of $N$.
Let notation $(x, y)$ indicate the greatest common divider (GCD) of $x$ and $y$. If $N$ is square-free, then we have
\begin{equation}
G(a,\chi ) = 0, \quad \forall (a,N) > 1 . \label{square-free}
\end{equation}
Conversely \cite{C07}, if $N$ is not square-free
\begin{equation}
G(a,\chi )=0, \quad \forall (a,N) = 1 .  \label{non-square-free}
\end{equation}
This remarkable fact suggests a dichotomy criterion for testing square-freeness, it represents the cornerstone of our algorithm.

We present the algorithm first for the relatively simple case where $N = p{q^2}$ ($p$, $q$ both prime)  and subsequently generalize it. The algorithm consists of two parts, as illustrated in Fig. 1. In the first part, we generate the state
\begin{equation}
\left| \phi  \right\rangle  = \frac{1}{{\sqrt {\varphi (N)} }}\sum\limits_{(m,N) = 1} {\chi (m,N)\left| m \right\rangle },
\label{phi}
\end{equation}
where the normalization coefficient ${\varphi (N)}$ represents Euler's function (number of integers smaller than $N$ that are coprime to $N$). van Dam and Seroussi \cite{WS} proposed a general method for preparing such a superposition state. They gave the example of computing the Legendre symbol, which is a special case of the Jacobi symbol, which reduces to the Legendre symbol when $N$ is prime. They also computed the Jacobi symbol for the case when the factorization of N is known. In our case, the factors of $N$ are not known. Thus we would adopt another technique for computing the Jaocbi symbol \cite{WSL}, which we discuss in the following.
The second part of the algorithm is to apply the quantum Fourier transform (QFT) to $\left| \phi  \right\rangle $.
The resulting state encodes the factors $p$ and $q$ of $N$, which can be retrieved by performing measurements on the qubits.

\begin{figure*}
\centering
\includegraphics[height=3cm]{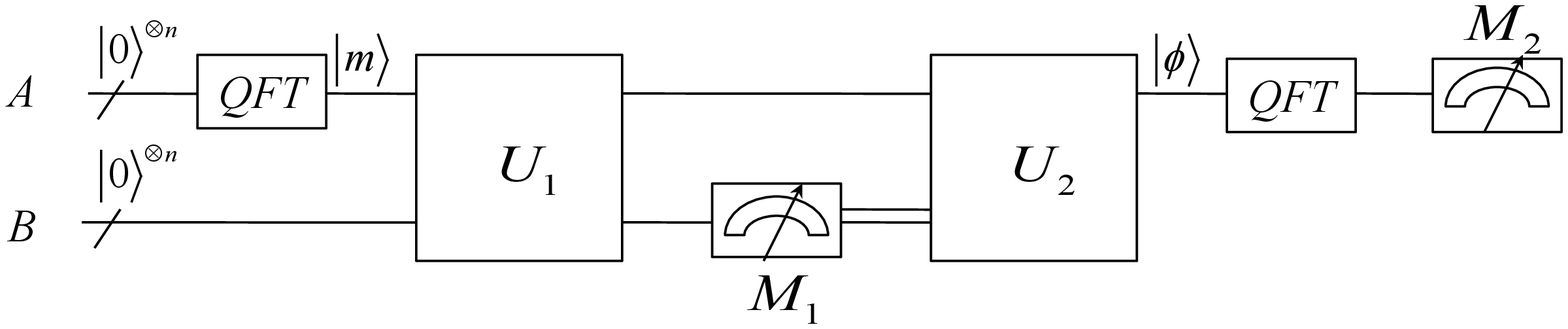}
\caption{\textbf{Outline of quantum circuit for computing the square-free part for $N=pq^2$.}
The procedure denoted as  $\Omega $ in the text consists of two main parts.
In the first part, we generate the state $\left| \phi  \right\rangle $; in the second part, we apply the quantum Fourier transform (QFT) to it.
Single lines represent qubits, and boxes represent operations.
Time runs from left to right. The transformation ${U_1}$ and ${U_2}$ are defined by Eq. (\ref{U1}) and Eq. (\ref{U2}). The meters $M_1$ and $M_2$ represent the measurements. The double lines coming from $M_1$ carry the classical bits, here the algorithm continues only if register $B$ collapses to $1$.}
\end{figure*}

Now we discuss the details of the algorithm.
Set $n = \left\lceil {\log N} \right\rceil $, the smallest integer for which $2^n \ge N$.
We need two main registers $A$ and $B$, both initialized to ${\left| 0 \right\rangle ^{ \otimes n}}$.
Additional registers needed for storing auxiliary variables and constants are not represented explicitly for simplicity.
The first part starts with a state uniformly superposed from $1$ to $N-1$, which is prepared just by an $N-1$ dimensional Fourier transform on register $A$ and a subsequent addition with $1$
\begin{equation}
\left| 0 \right\rangle _A^{ \otimes n}\left| 0 \right\rangle _B^{ \otimes n} \to \frac{1}{{\sqrt {N - 1} }}\sum\limits_{m = 1}^{N - 1} {{{\left| m \right\rangle }_A}} \left| 0 \right\rangle _B^{ \otimes n}.  \nonumber
\end{equation}
Note that this Fourier transform is of order $N-1$, and it was known \cite{MZ} that the quantum fast Fourier transform can be made exact for arbitrary orders. Next we compute the greatest common divisor of $m$ and $N$ into register $B$
\begin{equation}
{U_1} : \frac{1}{{\sqrt {N - 1} }}\sum\limits_{m = 1}^{N - 1} {{{\left| m \right\rangle }_A}} \left| 0 \right\rangle _B^{ \otimes n} \to \frac{1}{{\sqrt {N - 1} }}\sum\limits_{m = 1}^{N - 1} {{{\left| m \right\rangle }_A}{{\left| {(m,N)} \right\rangle }_B}}.  \label{U1}
\end{equation}
Classically, the GCD problem can be efficiently solved by the classical Euclidean algorithm in quadratic polynomial time. In order not to involve the complicated division arithmetics of the Euclidean algorithm, we prefer to adopt the extended Euclidean algorithm \cite{K}. The extended Euclidean algorithm can be directly generalized to a quantum GCD algorithm that operates on a superposition state with the same computational complexity (see supplementary information for the quantum network construction).

We then take a measurement $M_1$ of register $B$. If the result is not $1$, then it must be $p$ or $q$ or $q^2$ and clearly the algorithm already succeeds. However, it's highly possible that we would not obtain such results, and the algorithm continues. This is because the probability of obtaining $(m,N)=1$ is $\varphi (N)/(N-1) = (p - 1)(q - 1)/(pq-1)$, which asymptotically approaches $1$ for sufficiently large $p$ and $q$. If $M_1$ results in $1$, we get
\begin{equation}
\frac{1}{{\sqrt {\varphi (N)} }}\sum\limits_{(m,N) = 1} {{{\left| m \right\rangle }_A}{{\left| 1 \right\rangle }_B}}. \nonumber
\end{equation}
The next step is to obtain the state $\left| \phi  \right\rangle $ as given in (\ref{phi}), i.e., we do the following unitary operation on register $A$
\begin{equation}
{U_2}: \frac{1}{{\sqrt {\varphi (N)} }}\sum\limits_{(m,N) = 1} {{{\left| m \right\rangle }}}  \to \frac{1}{{\sqrt {\varphi (N)} }}\sum\limits_{(m,N) = 1} {\chi (m){{\left| m \right\rangle }}},
 \label{U2}
\end{equation}
where ${\chi (m)}$ are $1$ or $-1$ as by the definition of Jacobi symbol, and register $B$ is omitted. The key part of $U_2$ is to compute the Jacobi symbol $\chi (m)$ for all $(m,N)=1$. Classically, the Jacobi symbol can be efficiently solved by many algorithms. There exists \cite{SS} a binary algorithm which has the advantage of lower complexity and easier implementation on a binary computer. The binary algorithm can be seen as a variant of the extended Euclidean algorithm, and hence can also be extended to a quantum algorithm (see supplementary information for the quantum network construction).

\begin{figure*}
\centering
\includegraphics[height=10cm]{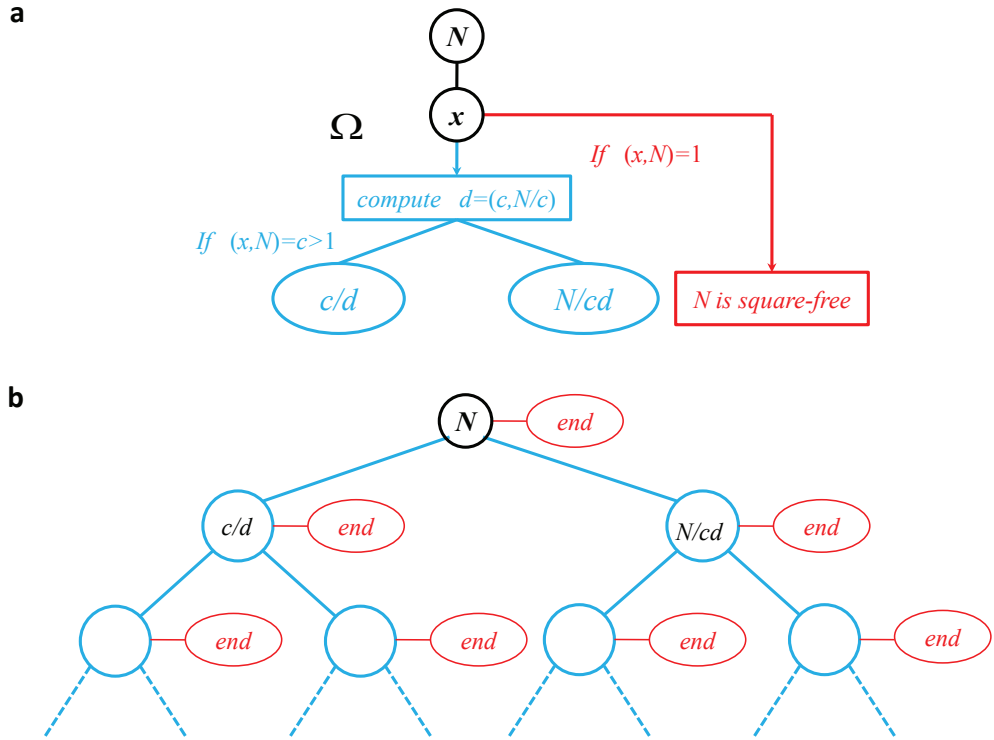}
\caption{\textbf{Schematic flow chart of the recursive quantum algorithm for computing the squrare-free part of an arbitrary odd integer $N$.}
\textbf{a}, Possible outcomes of applying the algorithm $\Omega $ on an arbitrary odd integer $N$: either return a factor $c$ or else ensure that $N$ is a square-free number with the square-free part $r = N$. If a factor $c >1$ is returned, and $N$ is tested to be not square-free, then the problem is converted to two smaller subproblems for $c/d$ and $N/cd$ where $d$ is the greatest common divisor of $c$ and $N/c$. This serves as the subroutine of the recursive quantum algorithm. \textbf{b}, Recursive algorithm for a general $N$. Different colors are used to designate two different outcomes after applying the subroutine $\Omega$.
The red color denotes that number is square-free, then this branch terminates. The blue color denotes the other outcome;
in this case, the algorithm proceeds to the next step of recursion.
The $\Omega$ operation needs to be performed at most $\log N$ times to solve this problem.}  \label{alg}
\end{figure*}

As the last step of the algorithm, we take a Fourier transform on $\left| \phi  \right\rangle $ and obtain
\begin{align}
 \left| \psi  \right\rangle   & = \frac{1}{{\sqrt {N \varphi (N) } }}\sum\limits_{k = 0}^{N - 1} {\left( {\sum\limits_{(m,N) = 1} {\chi (m){e^{2\pi imk/N}}} } \right)\left| k \right\rangle }  \nonumber \\
 & = \frac{1}{{\sqrt {N\varphi (N)} }}\sum\limits_{k = 0}^{N - 1} {G(k,\chi )\left| k \right\rangle }.
\end{align}
According to the properties  (\ref{square-free}) and  (\ref{non-square-free}) of the Gauss sum,
all amplitudes vanish unless $k$ shares a nontrivial common factor with $N$. If we perform a measurement $M_2$ on the register, it always collapses to a state $\vert k_0 \rangle$, whose GCD with $N$ is a non-trivial factor $p$ or $q$ of $N$. It therefore yields the complete decomposition of $N$.

We now determine the computational complexity of this algorithm. All the transformations involved in the algorithm, including the extended Euclidean algorithm for GCD and Jacobi symbol and QFT, require $O({(\log N)^2})$ elementary gate operations \cite{NC}. Thus this algorithm has only a polynomial-time complexity.

For a general $N$ with possibly many distinct prime factors (square-freeness of $N$ is unknown), the procedure outlined above may not work. However, it can be generalized to include this case, and the generalized algorithm remains simple and efficient. We refer to the algorithm described above as $\Omega$ and discuss now the generalized algorithm, which includes $\Omega$ as a subroutine.

As we discussed, the algorithm $\Omega$ includes two measurements, $M_1$ and $M_2$. With a certain probability, $M_1$ yields a nontrivial factor of $N$. If this does not happen, we proceed to the second measurement $M_2$. Two possibilities will occur at $M_2$ due to the dichotomy property of Gauss sum  (\ref{square-free}, \ref{non-square-free}) :
we obtain (\emph{i}) a non-trivial factor of $N$ if $N$ is not square-free, or (\emph{ii}) a result coprime to $N$, which signifies that $N$ is definitely square-free. As a result, no matter whether $\Omega$ ends at $M_1$ or $M_2$, it either yields a non-trivial factor (say $c$) of $N$ or determines that $N$ is square-free. In the latter case, we have succeeded already, hence the algorithm finishes. In the former case, if the two parts $c$ and $N/c$ share a common factor $d= (c, N/c)$, we know that $d^2$ is a factor of the square part $s^2$ of $N$.
We thus can split the problem of finding the square-free part $r$ of $N$ into two smaller problems:
finding the square-free parts of $c/d$ and $N/(cd)$.
From the solutions of  these subproblems, we find the corresponding parts of $N$ as
\begin{equation}
\begin{split}
r &= R(N) = R(c/d) \cdot R(N/cd) \\
s^2 &= S(N) = S(c/d) \cdot S(N/cd)\cdot{d^2} .
\end{split}
\end{equation}
Here, $R( \cdot )$ and $S( \cdot )$ represent the square-free part and the square part of their argument, respectively.
Clearly, this procedure can be iterated until all branches have determined that the arguments are square-free.
Figure 2 illustrates this recursive procedure.

The execution time of the extended algorithm reaches a maximum when each execution of $\Omega$ yields just one factor, but clearly, the number of repetitions is still bounded by $O(\log N)$. Each execution of the subroutine $\Omega$ requires at most $O({(\log N)^2})$ steps.
The worst-case complexity of the extended algorithm is therefore $O({(\log N)^3})$. Actually, we have a better estimation of how long it takes untill the algorithm succeeds. This is by virtue of the observation that $M_2$ yields the square part with high probability, and calculations show that the algorithm will finish with high probability in just $O({(\log N)^2}{(\log \log N)^2})$ (see methods).

Classically, finding the square-free part of an integer is known to be hard. It was argued \cite{OU} that the best method known for its solution is through factorization.
The fastest classical algorithm for factorization would  be the number field sieve \cite{CP}, which requires $O(\exp(c(\log N)^{1/3} ( \log \log N)^{2/3}))$ steps.
Thus the quantum algorithm presented here offers an exponential speed-up over the classical algorithm. A feasible alternative to our algorithm would be to use Shor's algorithm to obtain the complete decomposition of $N$ also in polynomial time. Application of Shor's algorithm yields, with some probability, two divisors of $N$ in time $O({(\log N)^2}\log \log N\log \log \log N)$ \cite{NC}. Like our algorithm, Shor's algorithm would thus also be applied repetitively, with the number of iterations bounded by $O(\log N)$.
The overall computational complexity using Shor's algorithm would be $O({(\log N)^3}\log \log N\log \log \log N)$.
Figure 3 compares the computational costs of the three algorithms described above, clearly showing the
increase in computational efficiency by the algorithm presented here.

\begin{figure}[h]
\centering
\includegraphics[height=12cm]{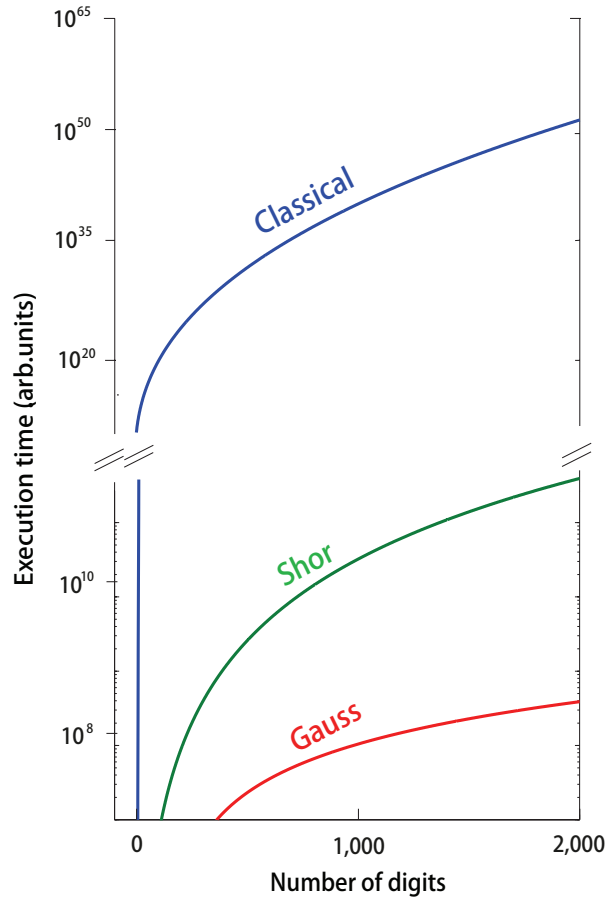}
\caption{\textbf{Comparison between the computational costs of the three algorithms discussed in the text.}
Both quantum algorithms offer exponential speedup over the classical methods.
For hundreds of digits, our algorithm is almost two orders of magnitude faster than the Shor's algorithm.}
\label{complexity}
\end{figure}

Our algorithm relies on the mathematical properties of the Gauss sums. The possibility of using the periodicity properties of Gauss sums for factorization was suggested earlier \cite{M, M06} and the feasibility of this approach was demonstrated in various physical systems including nuclear magnetic resoance \cite{M07, MRPS, PS}, cold atoms \cite{G} and superconducting circuits \cite{NN}. However, these schemes did not use the specific properties of quantum mechanical systems.
They can be implemented in classical as well as in quantum systems and the scaling properties are therefore not superior to other classical algorithms. In contrast, the algorithm that we have described in this paper relies on quantum superpositions and is both \emph{efficient and deterministic} in solving the square-free part computation problem, even demonstrates advantages over Shor's approach.
In Shor's algorithm, the major cost comes from the modular exponentiation operation, while Gauss sums can be generated through  $O({(\log N)^2})$ modular square operation. In our algorithm, we have noticed that Gauss sum evaluations are closely related to the factorization of $N$.
While we have not found such an algorithm so far, it may thus be possible to develop a quantum algorithm on the basis of Gauss sums that solves integer factorization.

\section*{Methods}
\subsection{Realization of $U_1$.}
$U_1$ is to compute the greatest common divisor of $m$ and $N$. Classically, the GCD problem can be efficiently solved based on the famous extended Euclidean algorithm. There is a variant of this algorithm, called the binary GCD algorithm, which can be more conveniently performed on a binary computer. We adopt this method here, and succeed in finding a quantum network that performs the binary GCD algorithm on a quantum superposition state (see supplementary information for details).

\subsection{Realization of $U_2$.}
In Fig. 1, the operation $U_2$ is realized through the following steps
\begin{equation*}
\begin{array}{lll}
(0). & \quad \sum\limits_{(m,N) = 1} {{{\left| m \right\rangle }_A}{{\left| 1 \right\rangle }_B}} & \mbox{initial state} \\
(1). &   \to \sum\limits_{(m,N) = 1} {{{\left| m \right\rangle }_A}{{\left| {\chi (m)} \right\rangle }_B}} & \mbox{apply the operation of Jacobi symbol computation}   \\
(2). &  \to \sum\limits_{(m,N) = 1} {{e^{i\pi \left( {\chi (m) - 1} \right)/2}}{{\left| m \right\rangle }_A}{{\left| {\chi (m)} \right\rangle }_B}} & \mbox{apply conditional phase shifts}   \\
(3). &   \to \sum\limits_{(m,N) = 1} {\chi (m){{\left| m \right\rangle }_A}{{\left| 1 \right\rangle }_B}}  & \mbox{apply step (1) in reverse order} \\
(4). &   ={\left| \phi  \right\rangle _A}{\left| 1 \right\rangle _B}, &
\end{array}
\end{equation*}
where we use the phase kickback trick \cite{WS} and the identity ${e^{i\pi \left( {\chi (m) - 1} \right)/2}} = \chi (m)$. The computation of Jacobi symbol can be implemented by binary Jaocbi algorithm (see supplementary information for details).

\subsection{Complexity estimation of the algorithm.}
In the following, we discuss the algorithm complexity for a general case $N$. To do this, we slightly change the algorithm presented in the text.
Our analysis is based on the finding: if at the measurement $M_2$ we obtain a result whose common divisor with $N$ is a square number,
then the common divisor must be the square part $s^2$ of $N$; and the probability of this case is larger than ${\left( {\varphi (N)/N} \right)^2}$ (see supplementary information for proofs).  Hence the algorithm can be altered in the way that if any branch of the algorithm proceeds to $M_2$ and results in a square number, then that branch terminates.

Denote $P( \cdot )$ as the probability of obtaining the square part of its argument by application of $\Omega$.  Let $p_k$ denotes the probability that the algorithm succeeds at the $k$-th iteration step. Obviously
\begin{equation*}{p_1} = P(N), \end{equation*}
where $P(N) \ge {\left( {\varphi (N)/N} \right)^2}$ .
If $\Omega$ does not succeed at the first step, and suppose we have obtained $c$ and $N/c$ and $d=(c,N/c)$, then
\begin{align}
{p_2} &= (1 - P(N))P\left( {\frac{c}{d}} \right)P\left( {\frac{N}{{cd}}} \right)  \nonumber \\
& \ge (1 - P(N)){\left( {\frac{{\varphi (c/d)}}{{c/d}}\frac{{\varphi (N/cd)}}{{N/cd}}} \right)^2} \nonumber \\
& \ge (1 - P(N)){\left( {\frac{{\varphi (N)}}{N}} \right)^2} \nonumber \\
& = (1 - P(N))P(N).  \nonumber
\end{align}
Here, the second inequality is valid because of a basic property of the Euler function
\begin{equation*}\varphi (mn) = \varphi (m)\varphi (n)\frac{{\gcd (m,n)}}{{\varphi (\gcd (m,n))}}.\end{equation*}
Analogously, we will have
\begin{equation*}{p_3} \ge {(1 - P(N))^2}P(N)\end{equation*}
\begin{equation*}...\end{equation*}
\begin{equation*}{p_k} \ge {(1 - P(N))^{k - 1}}P(N).\end{equation*}
Therefore, after $k$ steps, the probability that the algorithm still does not succeed is
\begin{equation*}
Q \le 1 - \sum\limits_{i = 1}^k {{{(1 - P(N))}^{i - 1}}P(N)}  = {(1 - P(N))^k} \le {\left( {1 - {{\left( {\frac{{\varphi (N)}}{N}} \right)}^2}} \right)^k}.
\end{equation*}
According to the inequality (Theorem 8.8.7 \cite{BS})
\begin{equation*}
\frac{{\varphi (N)}}{N} > \frac{1}{{{e^\gamma }\log \log N + \frac{3}{{\log \log N}}}},
\end{equation*}
where $\gamma = 0.5772...$ is the Euler-Mascheroni constant, and for a large $N$, $\varphi (N)/N > 1/(2\log \log N)$. So we have
\begin{equation}
Q < {\left( {1 - {{\left( {\frac{1}{{2\log \log N}}} \right)}^2}} \right)^k}.
\end{equation}
When $k = O({(\log \log N)^2})$, $Q \to 0$, this means, the algorithm doesn't need to go for $k = O(\log N)$ times, but would finish with high probability in $O({(\log \log N)^2})$ steps.

\section*{Acknowledgments}
This work was supported by the Chinese Academy Of Sciences and National Natural Science Foundation of China through grant no. 10975124,  and by the DFG through grant Su 192/19-1. The authors also thank W. Schleich who introduced us to the fascinating properties of Gauss sums.

\newpage
\renewcommand\figurename{\textbf{Supplementary Figure S \hspace{-8pt}}}
\setcounter{figure}{0}

\section*{Supplementary Information: An Efficient Deterministic Quantum Algorithm for the Integer Square-free Decomposition Problem  \\
Jun Li, Xinhua Peng, Jiangfeng Du, Dieter Suter }

\begin{appendix}

\section*{Gauss sum evaluation}
In the Gauss sum
\begin{equation}
G(a,\chi ) = \sum\limits_{m = 0}^{N - 1} {{\chi _N}(m){e^{2\pi iam/N}}} ,
\end{equation}
the  Jacobi symbol $\chi _N(m)$ is defined in terms of the Legendre symbols ${\chi _{{p_i}}}(m)$ with respect to the prime factors of $N$
\begin{equation}
{\chi _N}(m) = {\chi _{{p_1}}}{(m)^{{\alpha _1}}}{\chi _{{p_2}}}{(m)^{{\alpha _2}}} \cdots {\chi _{{p_k}}}{(m)^{{\alpha _k}}}.
\end{equation}
Here, the Legendre symbol ${\chi _{{p_i}}}(m)$ for a prime number $p_i$ is
\begin{equation}
{\chi _{{p_i}}}(m) = \left\{ {\begin{array}{*{20}{c}}
   {0} \hfill & {if \, m \equiv 0 (\bmod {p_i});} \hfill  \\
   {+1} \hfill & {if \, m \not \equiv 0 (\bmod {p_i}) \, \mathrm{ and } \, m = x^2 (\bmod {p_i}) }  \mbox{for some integer } x;  \hfill  \\
   {-1} \hfill & \mathrm{ otherwise. }\hfill  \\
\end{array}} \right.
\label{Legendre}
\end{equation}

If $N$ is square--free, the Jacobi symbol  is a primitive Dirichlet character $mod N$.
As shown in ref. \cite{C07}, the Gauss sum then has the
following important properties
\begin{equation}G(a,\chi ) = {\varepsilon _N}\chi (a)\sqrt N .
\label{GCond1}
\end{equation}
If, in addition, $(a,N) > 1$, by definition of the Jacobi symbol we'll have
\begin{equation}G(a,\chi ) = 0.\end{equation}

The general formula for the case that $N$ is not square--free
can be found in ref. \cite{C07} (p94, Exercise 12).
Here, we only need the special case where $(a,N)=1$.
Then, the general formula evaluates to
\begin{equation}G(a,\chi ) = 0.\end{equation}

\section*{Quantum Network for Extended Euclidean Algorithm}
In the main paper and methods, we point out that $U_1$ (greatest common divisor computation)
\begin{equation}
\frac{1}{{\sqrt N }}\sum\limits_{m = 0}^{N - 1} {\left| m \right\rangle \left| N \right\rangle }  \to \frac{1}{{\sqrt N }}\sum\limits_{m = 0}^{N - 1} {\left| m \right\rangle \left| {(m,N)} \right\rangle },
\end{equation}
and the key part of $U_2$ (Jacobi symbol computation)
\begin{equation}
\frac{1}{{\sqrt {\varphi (N)} }}\sum\limits_{(m,N) = 1} {\left| m \right\rangle \left| {\rm{1}} \right\rangle }  \to \frac{1}{{\sqrt {\varphi (N)} }}\sum\limits_{(m,N) = 1} {\left| m \right\rangle \left| {\chi (m)} \right\rangle } .
\end{equation}
can be efficiently solved based on the famous extended Euclid algorithm. There is a variant of this algorithm, called the binary extended Euclidean algorithm \cite{BS}, which can be more conveniently performed on a binary computer. The classical binary algorithms for GCD and Jacobi symbol are presented in supplementary Fig. 1 and supplementary Fig. 2 respectively.

We now try to construct a quantum version of the binary algorithms. The classical algorithm contains several conditional statements that translate into conditional control operations in the quantum algorithm. For some of them auxiliary registers are needed. Take binary GCD algorithm for example, corresponding to the conditional statements in supplementary Fig. 1, we add three additional registers called ``terminate or go on control register" (statement 7), ``even or odd control register" (statement 8, 9) and ``comparison control register" (statement 10). The three kinds of registers function as (1) ``terminate or go on": $0 \leftrightarrow v = 0$ and $1 \leftrightarrow v > 0$; (2) ``even or odd": $0 \leftrightarrow odd$ and $1 \leftrightarrow even$; (3)"comparison": $0 \leftrightarrow u < v$ and $1 \leftrightarrow u > v$ respectively. As illustrated in supplementary Fig. 3, the complete algorithm requires seven registers.
Details of the construction are illustrated in supplementary Fig. 5. Like in the classical case, the quantum circuit complexity for the GCD problem is of the order of ${\rm O}({(\log N)^2})$.

Analogously, the quantum network for Jacobi symbol computation can be constructed. Note that the conditional statements $5$, $9$, $12$ in supplementary Fig. 2 are easy to implement, because the modular properties of $u$ and $v$ with respect to $4$ and $8$ are only determined by their lowest digits. Besides, in the network, if $\chi (m) =-1$, we represent it by $r=N-1$. The explicit network for Jacobi symbol computation is illustrated in supplementary Fig. 4 and supplementary Fig. 6.

\section*{Complexity estimation of the algorithm}
In the complexity estimation of the algorithm (see methods), we made two arguments
\begin{itemize}
\item[\emph{1}.]
for an odd integer $N$, if $\Omega$ proceeds to $M_2$ with measurement result $o$ and $(o,N)$ is a square number, then $(o,N)$ must be the square part $s^2$ of $N$;
\item[\emph{2}.]
for an odd integer $N$, application of $\Omega$ will directly yield the square part of $N$ with probability larger than ${\left( {\varphi (N)/N} \right)^2}$.
\end{itemize}
Now we explain why the above arguments are valid.

\subsection{Proof of argument \emph{1}}
We now prove that at measurement $M_2$, if $(o,N)$ is a square number $z^2$, then it must be equal to $s^2$. To prove this, set $o=tz^2$ and $N=xz^2$.

We need to evaluate $G(tz^2, \chi)$. First it is rewritten as
\begin{equation}
G(t{z^2},\chi ) = \sum\limits_{m = 0}^{N - 1} {\chi (m){e^{2\pi it{z^2}m/N}}}  = \sum\limits_{(m,N) = 1} {\chi (m){e^{2\pi itm/x}}}.
\end{equation}
Let $m=kx+j$, where $0 \le k \le {z^2} - 1$ and $0 \le j \le x - 1$ such that $(kx+j,N)=1$. As
\begin{equation}
\chi (kx + j) = \left( {\frac{{kx + j}}{{x{z^2}}}} \right) = \left( {\frac{{kx + j}}{x}} \right) = {\chi _x}(j),
\end{equation}
there is
\begin{equation}
G(t{z^2},\chi ) = \sum\limits_{(j,N) = 1} {\left( {{\chi _x}(j){e^{2\pi itj/x}}\cdot\sum\limits_{(kx + j,N) = 1} 1 } \right)}.
\end{equation}
Since for a certain $j$
\begin{equation}
\sum\limits_{(kx + j,N) = 1} 1  = \frac{{\sum\limits_{(j,x) = 1} 1 \cdot\sum\limits_{(kx + j,N) = 1} 1 }}{{\sum\limits_{(j,x) = 1} 1 }} = \frac{{\varphi (N)}}{{\varphi (x)}},
\end{equation}
hence
\begin{equation}
G(t{z^2},\chi ) = \frac{{\varphi (N)}}{{\varphi (x)}}\sum\limits_{j = 0}^{x - 1} {{\chi _x}(j){e^{2\pi itj/x}}}  = \frac{{\varphi (N)}}{{\varphi (x)}}G(t,{\chi _x}).
\end{equation}
According to the Gauss sum formula, $G(t{z^2},\chi ) \ne 0$ if and only if $x$ is square-free.
Then as the square-free decomposition is unique, $x$ must be $r$ and $z^2$ must be $s^2$.

\subsection{Proof of argument \emph{2}}
For $o=ts^2$ and $(t,N)=1$, we then have
\begin{equation}
G(t{s^2},\chi ) = \frac{{\varphi (N)}}{{\varphi (r)}}\sum\limits_{j = 0}^{r - 1} {{\chi _r}(j){e^{2\pi itj/r}}} = \chi (t){\varepsilon _r}\sqrt r \frac{{\varphi (N)}}{{\varphi (r)}}.
\end{equation}
Now it is obvious that the probability of obtaining $o$ such that $o=ts^2$ and $(t,N)=1$ is
\begin{equation}
\frac{{\sum\nolimits_{(t,N) = 1} {{{\left| {G(t{s^2},\chi )} \right|}^2}} }}{{\varphi (N)N}}   =  \frac{{\varphi (r)r{\varphi ^2}(N)}}{{\varphi (N)N{\varphi ^2}(r)}}
 =  \frac{{\varphi (N)r}}{{N\varphi (r)}}.
\end{equation}
As the probability that the algorithm not terminate at measurement $M_1$ is ${\left( {\varphi (N)/N} \right)}$, the probability that the algorithm proceeds to measurement $M_2$ and gives the square part $s^2$ would be
\begin{equation}
\frac{r}{{\varphi (r)}}{\left( {\frac{{\varphi (N)}}{N}} \right)^2} \ge {\left( {\frac{{\varphi (N)}}{N}} \right)^2}
\end{equation}

\end{appendix}

\newpage
\begin{figure}
\centering
\includegraphics[width=10cm]{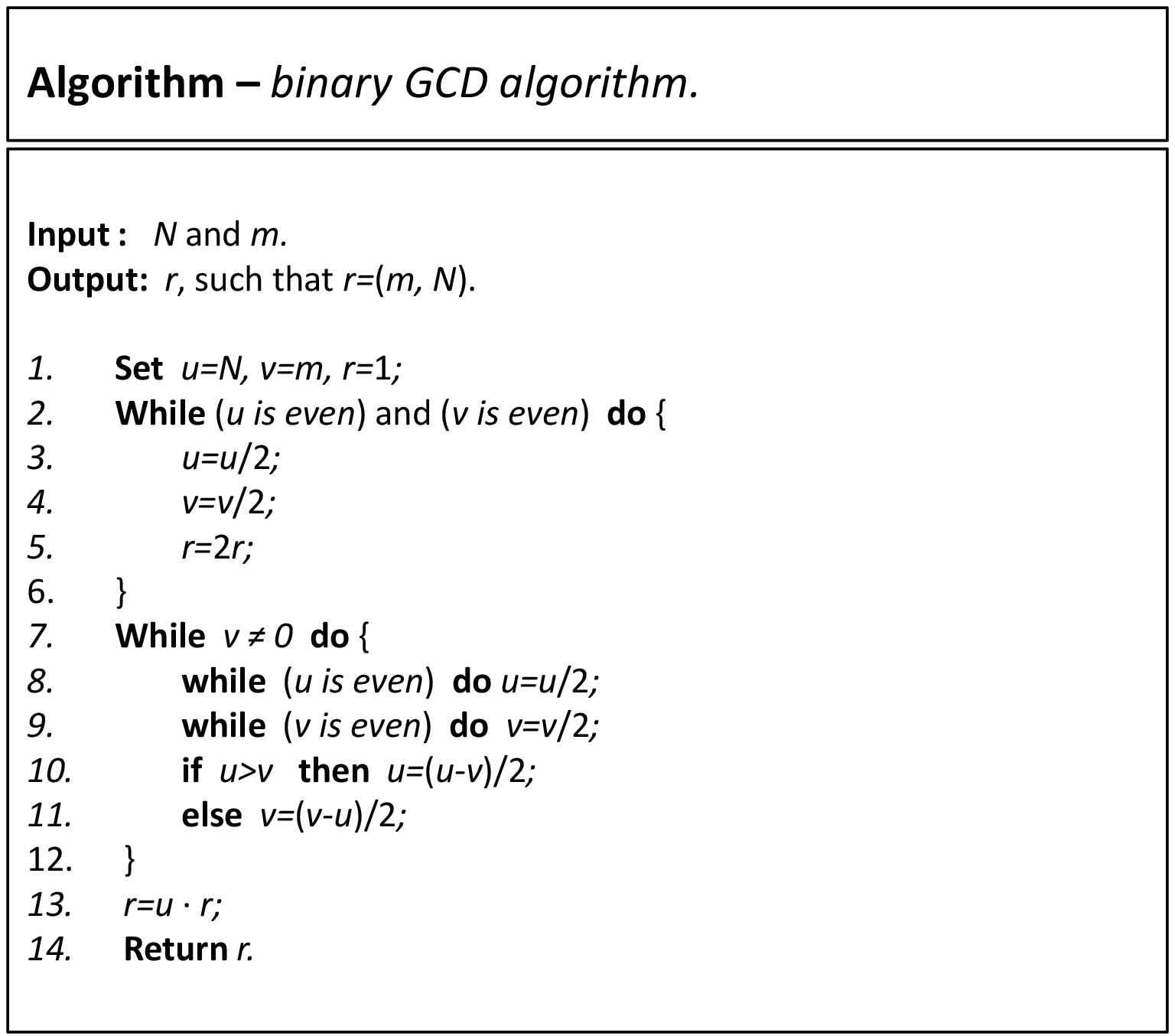}
\caption{Classical binary algorithm for computing GCD.}
\end{figure}

\begin{figure}
\centering
\includegraphics[width=10cm]{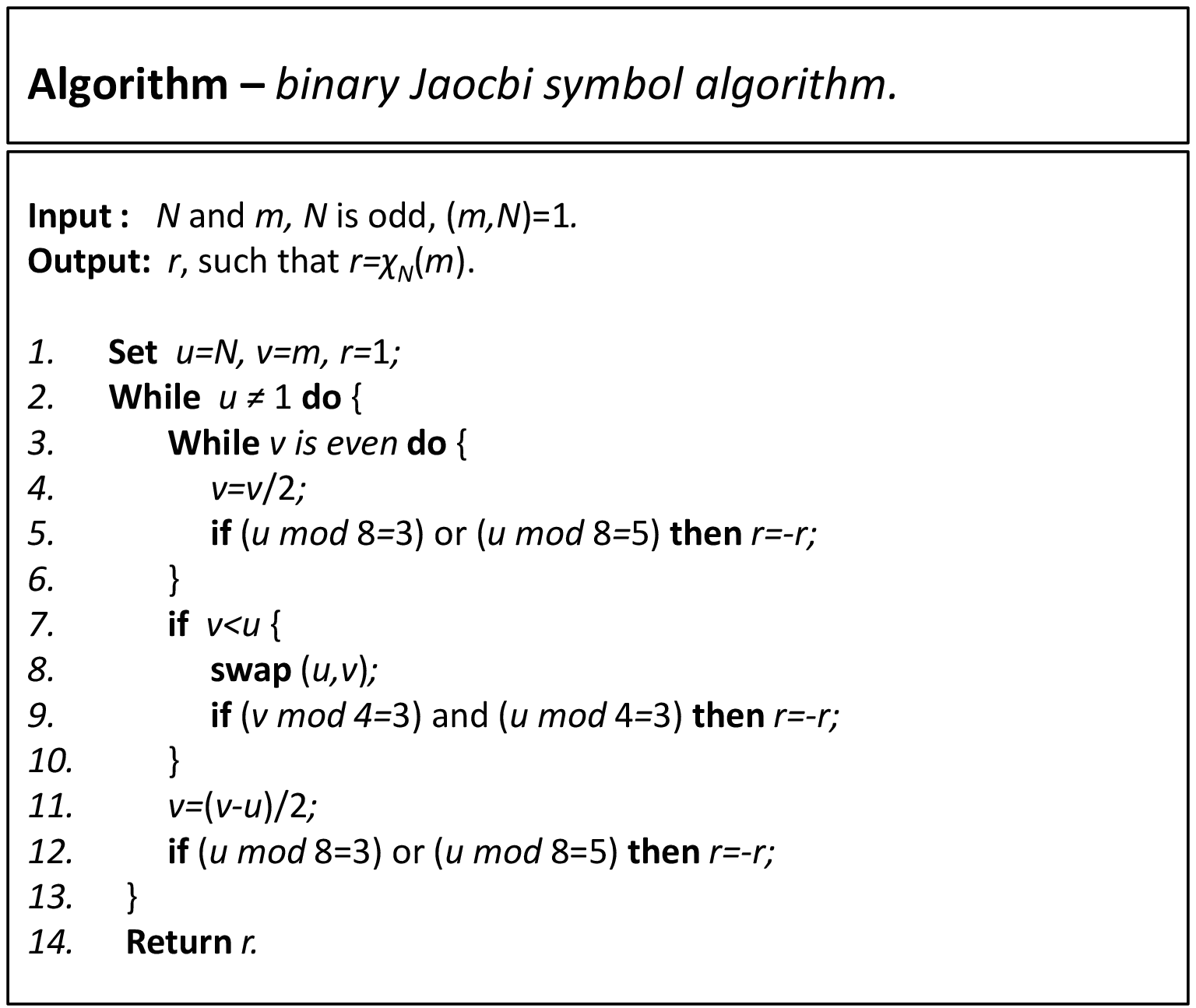}
\caption{Classical binary algorithm for computing Jacobi symbol.}
\end{figure}

\begin{figure}
\centering
\includegraphics[width=10cm]{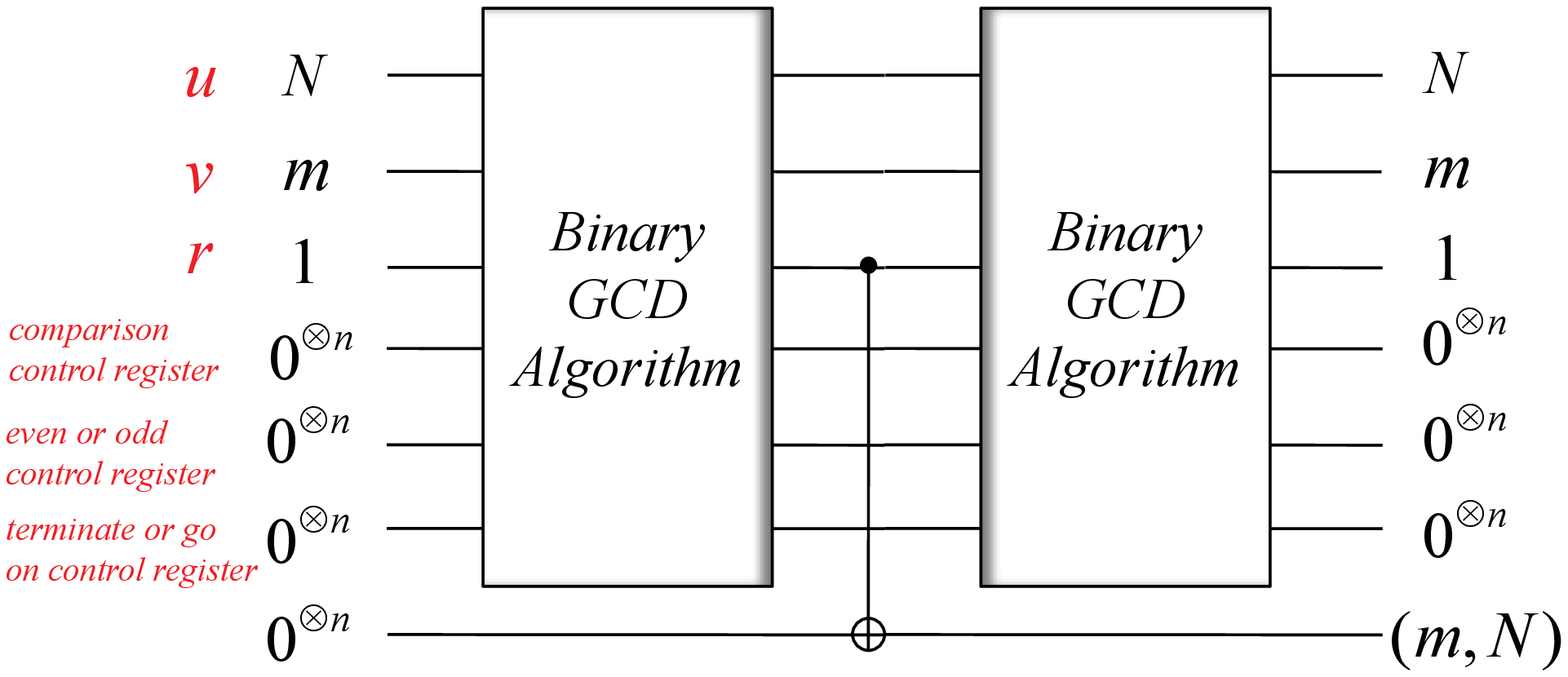}
\caption{\textbf{Outline of the circuit for binary GCD algorithm.} This represents the quantum generalization of the classical binary GCD algorithm. Each horizontal line represents a quantum register with $n$ qubits. The first three lines labeled $u,v,r$ represent the corresponding variables
from the classical algorithm, as represented in supplementary Fig. 1.}
\end{figure}

\begin{figure}
\centering
\includegraphics[width=10cm]{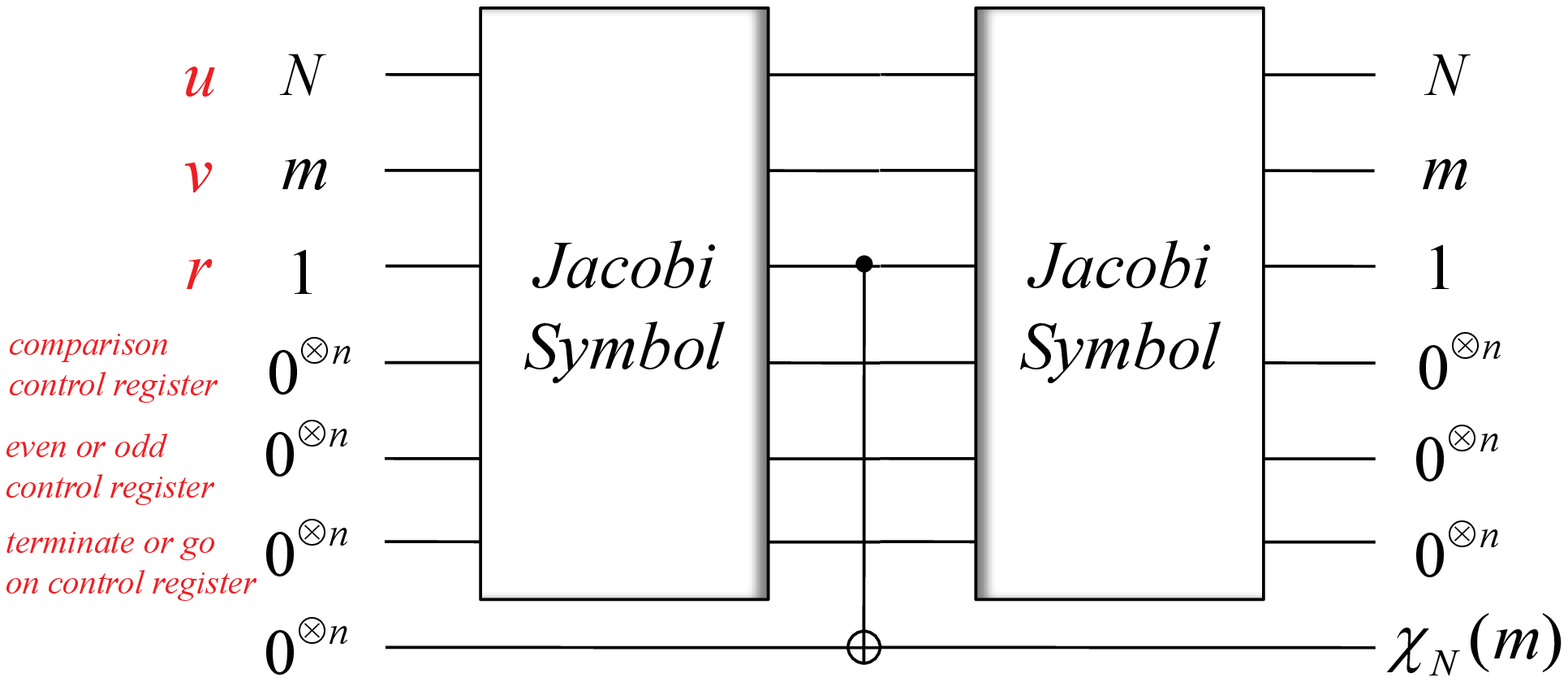}
\caption{\textbf{Outline of the circuit for binary Jacobi symbol algorithm.} This represents the quantum generalization of the classical binary Jacobi symbol algorithm. Each horizontal line represents a quantum register with $n$ qubits. The first three lines labeled $u,v,r$ represent the corresponding variables
from the classical algorithm, as represented in supplementary Fig. 2.}
\end{figure}

\begin{figure}
\centering
\includegraphics[width=16cm]{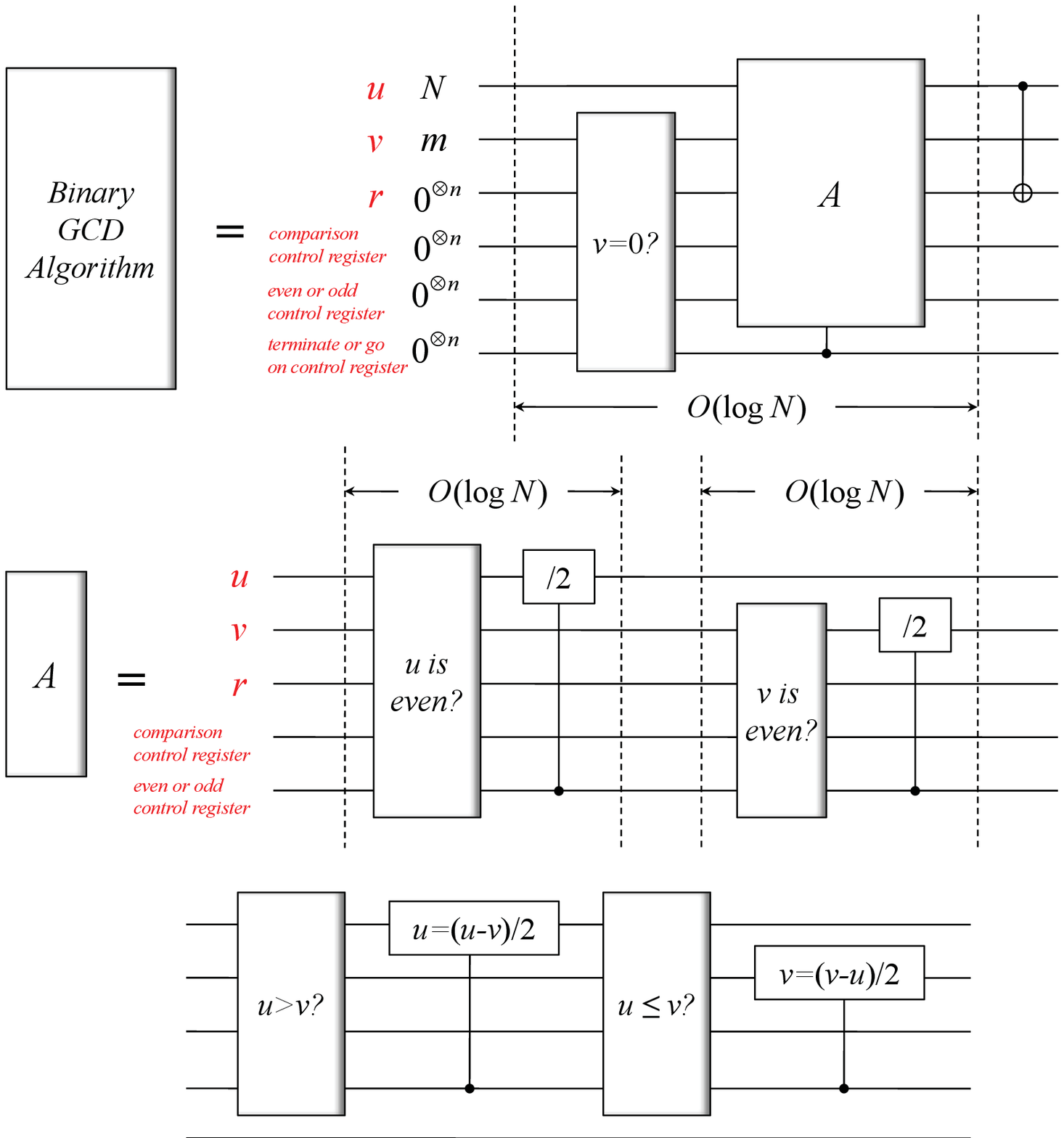}
\caption{\textbf{Quantum circuit for binary GCD algorithm.} The circuit is obtained through complete translation from the classical binary GCD algorithm.}
\end{figure}

\begin{figure}
\centering
\includegraphics[width=16cm]{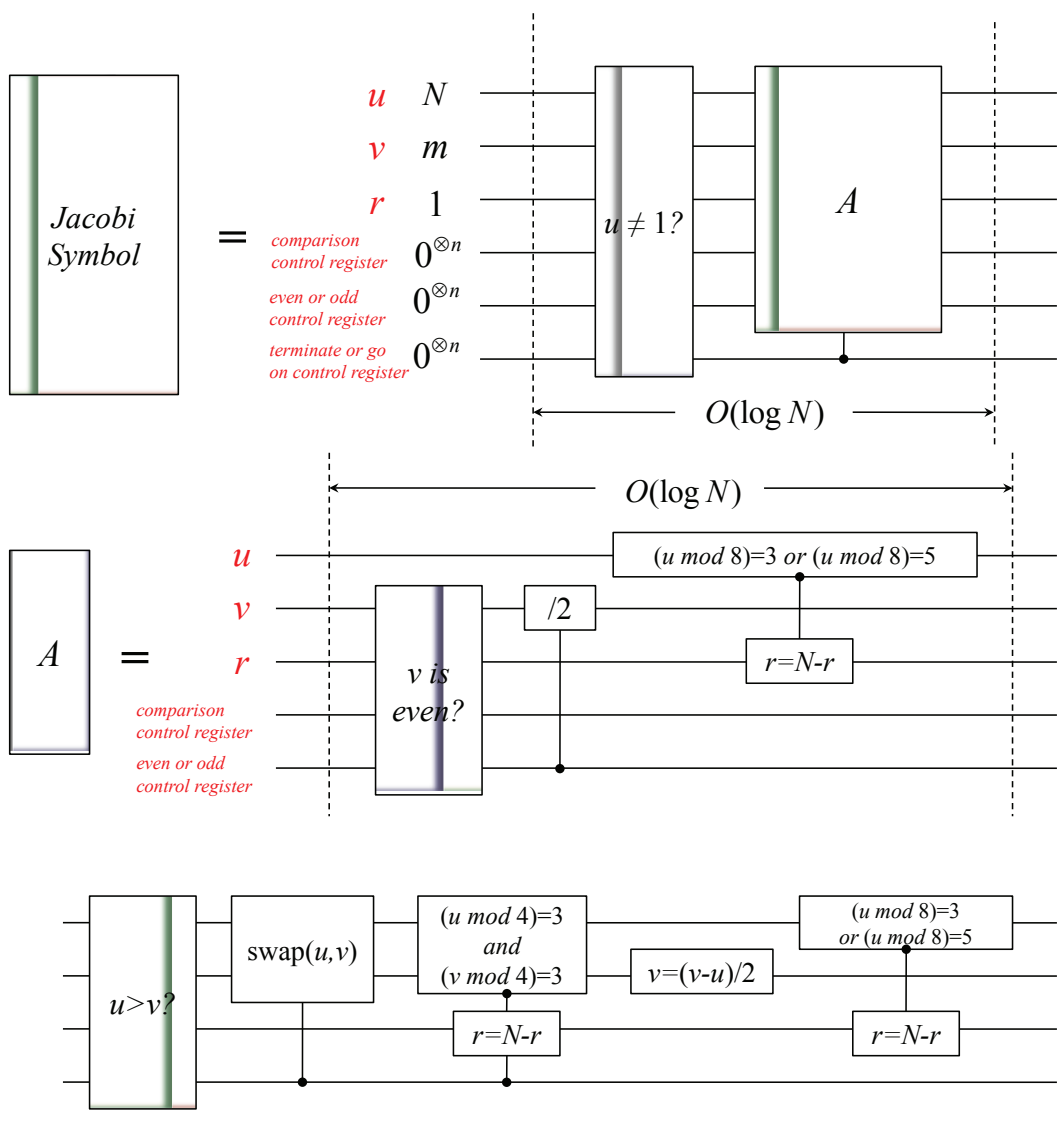}
\caption{\textbf{Quantum circuit for binary Jacobi symbol algorithm.} The circuit is obtained through complete translation from the classical binary Jacobi symbol algorithm.}
\end{figure}

\end{document}